\documentclass[10pt, reqno]{amsart}
\usepackage{amsmath, cite, amsthm, amssymb, amsfonts, mathrsfs}  
\usepackage[backref=page]{hyperref}
\usepackage{latexsym}
\usepackage{mathdots}
\usepackage{amssymb}
\usepackage{graphicx,bm,units,yfonts,yhmath,amscd}
\usepackage{verbatim}
\usepackage{enumerate}
\usepackage{lineno}
\newtheorem{theorem}{Theorem}[section]
\newtheorem{corollary}[theorem]{Corollary}
\newtheorem{lemma}[theorem]{Lemma}
\newtheorem{proposition}[theorem]{Proposition}
\theoremstyle{definition}
\newtheorem{example}[theorem]{Example}

\newcommand{\Ch}{\mathrm{Ch}}
\newcommand{\action}{T}

\begin{document}

\title[Non-commutative odd Chern numbers]{Non-commutative odd Chern numbers and  topological phases of disordered chiral systems}

\author{Emil Prodan}
\address{Physics Department, Yeshiva University, New York, NY, 10016, USA.}
\email{prodan@yu.edu}

\author{Hermann Schulz-Baldes}
\address{Department Mathematik, Universit\"{a}t Erlangen-N\"{u}rnberg, 91058 Erlangen, Germany}
\email{schuba@mi.uni-erlangen.de}

\thanks{E.~P. acknowledges support from the U.S. NSF grants DMS-1066045 and DMR-1056168; H.~S.-B. of the DFG}

\begin{abstract}
An index theorem for higher Chern characters of odd Fredholm modules over crossed product algebras is proved, together with a local formula for the associated cyclic cocycle. The result generalizes the classic Noether-Gohberg-Krein index theorem, which in its simplest form states that the winding number of a complex-valued function  over the circle is equal to the index of the associated Toeplitz operator. When applied to the non-commutative Brillouin zone, this generalization allows to define topological invariants for all condensed matter phases from the chiral unitary (or AIII-symmetry) class in the presence of strong disorder and magnetic fields, whenever the Fermi level lies in a region of Anderson localized spectrum. 
\end{abstract}

\maketitle


\section{Introduction}

For a classical compact manifold $\mathcal M$ of odd dimension $d$, the odd Chern character pairs with the $K_1$ group of classes of homotopically equivalent invertible matrices defined over $\mathcal M$, like the even Chern character pairs with the $K_0$ group of classes of homotopically equivalent idempotents defined over a compact manifold of even dimension \cite{ParkBook2008fh}. In the odd-dimensional case, the result of the pairing is the odd Chern number:
\begin{equation}
\label{OddChernNr1}
\Ch_d (U)
\;=\;
\frac{(\frac{1}{2}(d-1))!}{d !}
\;
\left(\frac{\imath}{2\pi}\right)^{\frac{d+1}{2}}\,
 \int\limits_{\mathcal M}\mathrm{Tr}\left(\big ( U^{-1} {\bf d} U \big )^d \right),
\end{equation}
which assigns an integer value to the homotopy class $[U]$ of a smooth function $U$ on $\mathcal M$ with values in the invertible matrices. This topological invariant is often referred to as generalized winding number because the case $d=1$ is precisely the winding number. In this case, the Noether-Gohberg-Krein index theorem \cite{No,GK} states that the integer $\Ch_1 (U)$ is equal to the index of the Toeplitz operator associated to $U$. The main mathematical result of this paper is to prove such an index theorem also for the case of odd dimension $d>1$, and for an adequate generalization of $\Ch_d (U)$ for $U$ lying in a crossed product algebra. This is to be seen as yet another situation where Connes' program of non-commutative geometry \cite{Con2} can be carried out. More precisely, let $(\Omega,T,\mathbb{Z}^d,\mathbb P)$ be a C$^*$-dynamical system given by a compact topological space $\Omega$ furnished with an action $T$ of $\mathbb{Z}^d$ and an invariant and ergodic probability measure $\mathbb P$. Associated to it is the reduced crossed product algebra of families of covariant operators under a projective representation of $\mathbb Z^d$ (see Section~\ref{Algebras} for details). Given a family of covariant, unitary operators $U=\{U_\omega\}_{\omega \in \Omega}$ on $\ell^2(\mathbb{Z}^d,\mathbb{C}^N)$ satisfying the condition
\begin{equation}
\label{Co}
\int_\Omega \mathbb P(d\omega) \,|\langle \bm x,\alpha|U_\omega|\bm y,\beta\rangle| \;\leq\; A\, e^{-\lambda |\bm x - \bm y|}\;, 
\qquad \bm x,\bm y\in\mathbb{Z}^d
\;,
\end{equation}
uniformly for some $A,\lambda>0$, the odd Chern number can be defined by
\begin{equation}
\label{OddChernNr1bis}
\Ch_d (U)
\;=\;
\frac{\imath\,(\imath \pi)^\frac{d-1}{2}}{d!!}\;  \sum_{\rho\in \mathcal S_d} (-1)^\rho \,\int_\Omega {\mathbb P} (d\omega)\;\mbox{\rm tr}_N\; 
\Big ( \langle \bm 0|\prod_{j=1}^d \big (\imath \, U_\omega^\ast  [U_\omega , X_{\rho_j}]  \big)|\bm 0\rangle \Big)
\;,
\end{equation}
where the sum runs over all permutations $\rho$ in the symmetric group $\mathcal S_d$ and $(-1)^\rho$ denotes their signature, $|\bm 0\rangle:\ell^2(\mathbb Z^d)\otimes\mathbb C^N\to\mathbb C^N$ is the partial isometry to all states at the origin $\bm 0\in\mathbb Z^d$, with adjoint $\langle\bm 0|$, and $\mbox{\rm tr}_N$ is the trace over the fiber $\mathbb C^N$, and finally $X_j$, $j=1,\ldots,d$, are the components of the position operator on $\ell^2(\mathbb{Z}^d,\mathbb{C}^N)$ defined by $(X_j \bm \psi)(\bm x) = x_j \bm \psi(\bm x)$ where ${\bm x}=(x_1,\ldots,x_d)\in\mathbb Z^d$. For periodic systems, \eqref{OddChernNr1bis} is the Fourier transform of \eqref{OddChernNr1} with $\mathcal M$ being the $d$-dimensional torus. Note that $\ell^2(\mathbb{Z}^d,\mathbb{C}^N) \simeq \mathbb C^N \otimes \ell^2(\mathbb Z^d)$ and both representations will be used.

\vspace{.1cm}

For the definition of the associated Toeplitz operator, let be given an irreducible representation of the complex Clifford algebra $Cl_{d}$ on $\mathrm{Cliff}(d)=\mathbb C^{d'}$ with $d'={2^{\frac{d-1}{2}}}$, provided by self-adjoint $\sigma_1,\ldots,\sigma_{d}$ satisfying $[\sigma_i,\sigma_j]=2\delta_{i,j}$. Then let us introduce the Dirac operator $D = \sum_{j=1}^d \sigma_j \otimes (X_j + x_i) $, shifted by $\bm x \in (0,1)^d$ to avoid zero modes and acting on the augmented Hilbert space $\mathrm{Cliff}(d) \otimes \ell^2(\mathbb{Z}^d,\mathbb{C}^N)$. Its phase is $F= D/|D|$ and the associated Hardy projection is $E=\frac{1}{2}(F+{\bf 1})$. The operator ${\bm 1} \otimes U_\omega$ on $\mathrm{Cliff}(d) \otimes \ell^2(\mathbb{Z}^d,\mathbb{C}^N)$ will also be denoted by $U_\omega$.

\begin{theorem}
\label{TheoIntro}
Let $d$ be odd and $U=\{U_\omega\}_{\omega \in \Omega}$ a covariant family of unitary operators on $\ell^2(\mathbb{Z}^d,\mathbb{C}^N)$ satisfying {\rm \eqref{Co}}. Then $E\, U_\omega E$ is $\mathbb P$-almost surely a Fredholm operator on $E\,\big (\mathrm{Cliff}(d) \otimes \ell^2(\mathbb{Z}^d,\mathbb{C}^N)\big )$ with $\mathbb P$-almost surely constant index given by
$$
\mathrm{Ind}(E\, U_\omega E)\; =\; \Ch_d (U)
\;.
$$
\end{theorem}

\vspace{.1cm}

Our main motivation to prove Theorem~\ref{TheoIntro} roots in its application to toplogical insulators. Solid state physicists are more accustomed with the even Chern numbers in dimension $d=2$ as they play a central role in the theory of the integer quantum Hall effect \cite{BELLISSARD:1994xj}. More recently, higher even Chern numbers have entered the theory of topological insulators \cite{Qi:2008cg,EssinPRL2009bv,LeungJPA2012er} and an index theorem similar to Theorem~\ref{TheoIntro} has been proved \cite{ProdanJPA2013hg}. Also the odd Chern numbers have already appeared in the literature on topological insulators from the chiral unitry class (also called the AIII-symmetry class) \cite{SchnyderPRB2008qy,Ryu:2010tq,MSH}. In fact, the ground state of a {\it periodic} system from this class can be uniquely characterized by a particular unitary matrix defined in the momentum-space (see Eq.~\eqref{eq-Qblock} below) and thus phases of such systems can be labelled by the odd Chern number \eqref{OddChernNr1}. In \cite{MSH}, the position-space formula \eqref{OddChernNr1bis}  was proposed as a phase label for disordered  systems in the chiral unitary class. Furthermore, for explicit 1 and 3-dimensional topological models from the chiral unitary class, the invariant was evaluated numerically in the presence of strong disorder \cite{MSH,SFP}, by implementing techniques from \cite{ProdanAMRX2013bn}. It was found that $\Ch_d (U)$ remains quantized and non-fluctuating as the disorder strength is increased, up to a critical disorder strength where the localization length of the system diverges and the invariant sharply changes its value. By Theorem~\ref{TheoIntro}, these findings are now analytically confirmed. The connection is being made by adapting Bellissard's  formalism \cite{BellissardLNP1986jf} describing homogeneous solid state systems by suited crossed product algebras. The most important conclusions drawn from this work are:

\begin{itemize}

\item[(i)] The ground state (represented by the Fermi projection) of any short-range homogeneous disordered quantum lattice system described by a Hamiltonian  from the chiral unitary class can be uniquely characterized by a canonical family of covariant unitary matrices $U_F=\{U_\omega\}_{\omega \in \Omega}$. In analogy with the Fermi projection, these operators will be called the Fermi unitary operators.

\item[(ii)] This family $U_F=\{U_\omega\}_{\omega \in \Omega}$ satisfies Eq.~\eqref{Co} whenever the Fermi level lies in a region of Anderson localization and therefore the non-commutative odd Chern numbers $\Ch_d(U)$ allow to distinguish different phases of chiral unitary systems. Here the localization regime is synonymous with the Aizenman-Molchanov bound on the fractional-powers of the Green's function \cite{Aizenmann1993uf}, which can indeed be proved to hold in a small and intermediate disorder regime for relevant models \cite{DDS}.

\item[(iii)] During continuous deformations of the Hamiltonian, the odd Chern number remains pinned at a quantized value as long as the Fermi level is located in a region of Anderson localization. At a phase transition, the localization length at the Fermi level has to diverge. 

\end{itemize}


These findings parallel the classification of topological solid state systems from the unitary class with no symmetry at all (also called class A), which is the only other class not requiring the use of a real structure. The prime example of a non-trivial topological phase in the unitary class is the  two-dimensional quantum Hall effect \cite{BELLISSARD:1994xj}. Higher even Chern numbers needed for the classification of systems in the unitary class in even dimension $d\geq 4$ were studied in \cite{ProdanJPA2013hg}. There are, however, a number of differences between A and AIII classes that ought to be stressed. First, the non-trivial topological phases in class A occur only in even space dimensions while in class AIII they occur only in odd space dimensions. And while the even Chern number in even dimension results from a pairing of the $K_0$ element specified by the Fermi projection with an even Fredholm module, hence leading to an even index theorem, the odd Chern number in odd dimension stems from the pairing of a $K_1$ element, the Fermi unitary operator $U_F$, with an adequate odd Fredholm module, hence providing an odd index theorem. From a physical point of view, there are also major differences between the unitary and the chiral unitary classes, which were already revealed by the numerics of \cite{MSH}.  A non-trivial topological phase from class A necessarily posses extended states at some energies above and below the Fermi level, because the even Chern number of the Fermi projection vanishes when the Fermi level is sent into the low and high energy limits (hence a topological transition must occur in the process). On the other hand, in chiral unitary systems a non-trivial topological invariant can go along with Anderson localization for all energies. Furthermore, the scenarios of phase transitions are quite distinct. In class A systems, two extended spectral regions above and below the Fermi level migrate towards each other until they collide and annihilate (as for Landau levels in quantum Hall systems). In the chiral unitary class, extended states can appear only at the Fermi level $E=0$ and only when crossing the boundary between topological phases. 

\vspace{.1cm}

The paper is organized as follows. Section~\ref{AIIIClass} presents the chiral unitary class of homogeneous aperiodic systems and shows how Theorem~\ref{TheoIntro} can be used for their topological classification. Section~\ref{Algebras} formalizes the algebra of covariant observables and introduces the non-commutative calculus. A smooth algebra (in the sense of \cite{Ren}) is defined together with a class of non-commutative Sobolev spaces. Section~\ref{ACProgram} implements Connes' program by defining a natural family of odd finitely summable Fredholm modules over the smooth algebra and by deriving a local index formula for its Chern character. Since the smooth algebra is involved here, this section covers the physics of disordered topological insulators, but only under the spectral gap assumption, {\it i.e.} when the Fermi level is located in a region empty of spectrum. This is relaxed in Section~\ref{Sobolev} to the physically more interesting assumption of a mobility gap, {\it i.e.} when the Fermi level is located in a region of Anderson localized spectrum. This is achieved by extending the local index formula over the non-commutative Sobolev space mentioned above, hence enabling us to formulate Theorem~\ref{TheoIntro} under condition \eqref{Co}. This is the central result of the paper. Section~\ref{Sobolev} also provides the mathematical statements supporting the claims (i), (ii) and (iii) above. 

\section{Topological classification of homogeneous chiral unitary systems}
\label{AIIIClass}

A quantum system described by a Hamiltonian $H$ on $\mathbb C^{2N} \otimes\ell^2(\mathbb Z^d)$ is said to have a chiral symmetry if for some involutive unitary $J\otimes \bm 1$, denoted simply by $J$,
\begin{equation}
\label{eq-ChiralHamiltonian}
JHJ\;=\;-\,H
\;,
\qquad
J^*J\;=\;J^2={\bf 1}\;.
\end{equation}
The system described by such $H$ is then said to be in the chiral unitary class (or AIII class of the Cartan classification). The chiral orthogonal and chiral symplectic classes are defined by a further symmetry the definition of which requires a real structure on the Hilbert space, but this will not be considered here. The unitary $J$ has eigenvalues $-1$ and $1$ which are supposed to have equal multiplicity. We will always work in the spectral representation of $J$ so that it can be assumed to be of the block form
\begin{equation}
\label{eq-Sblock}
{J}
\;=\;
\begin{pmatrix}
{\bf 1} & 0
\\
0 & -{\bf 1}
\end{pmatrix}
\;.
\end{equation}
This induces a grading of the Hilbert space, with summands which are equal $\mathbb C^N \otimes \ell^2(\mathbb Z^d)$. In this grading, a Hamiltonian with chiral symmetry \eqref{eq-ChiralHamiltonian} is of the form
\begin{equation}
\label{eq-Hblock}
H
\;=\;
\begin{pmatrix}
0 & A^*
\\
A & 0
\end{pmatrix}
\;,
\end{equation}
with $A$ being an operator on $\mathbb C^N \otimes \ell^2(\mathbb Z^d)$. An immediate consequence of the chiral symmetry is that the energy spectrum is invariant under the reflection $E\leftrightarrow -E$. For a variety of physical reasons, the Fermi level is always fixed to be the reflection symmetric point $E=0$. Hence the Fermi projection is $P_F=\chi(H< 0)$, where $\chi$ denotes the characteristic function. The sign function $Q={\rm sgn}(H)$ of the Hamiltonian is often called the flat band version of $H$. It is an odd function of $H$ and therefore it also satisfies the chiral symmetry $JQJ=-Q$. Consequently, it is also of the block form
\begin{equation}
\label{eq-Qblock}
Q
\;=\;
\begin{pmatrix}
0 & U_F^*
\\
U_F & 0
\end{pmatrix}
\;.
\end{equation}
It will be assumed that $E=0$ is not an eigenvalue of $H$ so that $Q^2={\bf 1}$. This then implies that $U_F$ is a unitary operator on $\mathbb C^N \otimes \ell^2(\mathbb Z^d)$ called the Fermi unitary operator. Alternatively, $U_F$ is the unitary of the polar decomposition of $A$ in \eqref{eq-Hblock}. Clearly $U_F$ determines the Fermi projection, and vice versa. Hence invariants of the ground state, represented by $P_F$ in the single particle picture adopted here, can be defined in terms of $U_F$. This will be done in the sequel under adequate hypothesis on $H$, by appealing to Theorem~\ref{TheoIntro}.

\subsection{Periodic chiral systems} 

The generic translation invariant lattice Hamiltonian on $\mathbb C^{2N} \otimes \ell^2(\mathbb Z^d)$ is of the form
\begin{equation}
\label{LatticeH0}
H
\; =\;
\sum_{{\bm a}\in \mathbb Z^d} t_{\bm a}\otimes S_{\bm a}\;,
\qquad
{\bm x}\in\mathbb Z^d
\;,
\end{equation}
where  each ${t_{\bm a}}$ is a $2N \times 2N$ matrix with complex entries such that $t_{-\bm a}=(t_{\bm a})^*$, and $S_{\bm a}$ are the lattice shifts by $\bm a$, $S_{\bm a}|\bm x,n \rangle = |\bm x + \bm a,n \rangle$ with $n=1,\ldots,2N$. 
Throughout, a finite hopping-range condition will be assumed to hold, namely that $t_{\bm a}$ is non-zero only for a finite number of $\bm a\in \mathcal R \subset \mathbb Z^d$. The chiral symmetry of $H$ is guaranteed if and only if $J\,t_{\bm a}J=-\,t_{\bm a}$. Due to translational symmetry, the discrete Fourier transform $\mathcal F:\ell^2(\mathbb Z^d,\mathbb C^{2N})\to L^2(\mathbb{T}^d, \mathbb{C}^{2N})$ leads to a direct integral representation of $H$, namely $\mathcal F H\mathcal F^*=\int^\oplus d{\bm k}\,H({\bm k})$, with Bloch Hamiltonians given by $2N\times 2N$ matrices
$$
H({\bm k}): \mathbb{C}^{2N} \rightarrow \mathbb{C}^{2N}\;, 
\qquad 
H({\bm k}) \;=\; \sum_{\bm a\in \mathcal R} t_{\bm a} \,e^{\imath {\bm a}\cdot {\bm k}}
\;.
$$
If the system is an insulator, zero energy $0$ lies in a gap of $H$. Then the Fermi projection $P_F$ and the flat band Hamiltonian $Q$ have similar direct integral representation. As the chiral symmetry $J \otimes {\bf 1}$ acts only on the fiber, both $H({\bm k})$ and $Q({\bm k})$ are off-diagonal block matrices, similar as in \eqref{eq-Hblock} and \eqref{eq-Qblock} respectively. The off diagonal blocks of $Q({\bm k})$ are $N\times N$ unitary matrices $U_F({\bm k})$ depending analytically on ${\bm k}$. For these unitaries the classical odd Chern number can be defined by \eqref{OddChernNr1}, with $\mathcal M=\mathbb T^d$. That this provides an integer that allows to distinguish periodic systems in the chiral unitary class was proposed in the pioneering works \cite{SchnyderPRB2008qy,Ryu:2010tq}, where it was referred to as the generalized winding number. As will be seen shortly, this can be generalized to aperiodic systems. Before going on, let us provide a simple periodic model in dimension $d=1$ having a non-trivial odd Chern number and thus a topological phase.

\begin{example}
In \cite{MSH} the Hamiltonian
\begin{equation}
\label{Model1}
H
\;=\;
\tfrac{1}{2}(\sigma_1+\imath \sigma_2)\otimes S_1 + \tfrac{1}{2}(\sigma_1-\imath \sigma_2)\otimes S_{-1}+m \, \sigma_2 \otimes \bm 1
\;,
\end{equation}
acting on $\mathbb C^2 \otimes \ell^2(\mathbb Z)$ was considered. Here $\sigma_1,\sigma_2,\sigma_3$ are the Pauli matrices. It has the chiral symmetry \eqref{eq-ChiralHamiltonian} with $J=\sigma_3 \otimes \bm 1$.  The odd Chern number $\Ch_{1}(U_F)$ reduces to the classical winding number and can be computed to be $\Ch_{1}(U_F) = -1$ for $m\in (-1,1)$, and $\Ch_{1}(U_F) = 0$ otherwise. The spectral gap of the model closes exactly at $m=\pm 1$ where the invariant switches between quantized values.
\hfill $\diamond$
\end{example}

\subsection{Homogeneous chiral systems} 
\label{sec-HomSys}

The periodicity of a one-particle model may be broken by a magnetic field, a random aperiodic perturbation or both. Typically one then has not only one tight-binding Hamiltonian on $\mathbb C^{2N} \otimes \ell^2(\mathbb Z^d)$, but rather a family $\{H_\omega\}_{\omega\in\Omega}$ indexed by a parameter from a compact probability space $(\Omega,{\mathbb P})$ of disorder configurations. These configurations can be shifted in physical space so that there is a group action  $\action$ of $\mathbb Z^d$ on $\Omega$ by homeomophisms. The probability measure ${\mathbb P}$ is supposed to be invariant and ergodic w.r.t. $\action$, so that indeed all the ingredients of a C$^*$-dynamical system are given. The Hamiltonians are then of the form
\begin{equation}
\label{LatticeHamOmega}
H_\omega
\;=\;
\sum_{\bm a \in \mathcal R}\sum_{\bm x \in \mathbb Z^d}  e^{\imath {\bm a}\wedge {\bm x}} \, t_{\bm a}(T_x \omega) \otimes |\bm x \rangle \langle \bm x | S_{\bm a}
\;,
\end{equation}
where $\wedge$ is an anti-symmetric bilinear form incorporating the effect of a constant magnetic field by means of a Peierls phase-factor and the $2N \times 2N$ hopping matrices $t_{\bm a}$ are now continuous functions of $\omega$, {\it i.e.} they belong to $M_{2N} \otimes C(\Omega)$ with $C(\Omega)$ as usual endowed with the topology induced by the sup-norm $\|\phi\|_{C(\Omega)} = \sup_{\omega \in \Omega} |\phi(\omega)|$. Still $\mathcal R\subset\mathbb Z^d$ is assumed to be finite. The collection $\{H_\omega\}_{\omega \in \Omega}$ of Hamiltonians defines a covariant family of operators, in the sense that
\begin{equation}
\label{eq-covrel}
V_{\bm a} H_\omega V_{\bm a}^\ast\;=\;H_{\action_{\bm a}\omega}
\;,
\end{equation}
for the magnetic translations defined by
\begin{equation}
\label{eq-Vdef}
V_{\bm a} \;=\;\bm 1 \otimes e^{-\imath {\bm a} \wedge {\bm X}} S_{\bm a}
\;.
\end{equation}
Reciprocally, any covariant family of finite range Hamiltonians takes the form \eqref{LatticeHamOmega}. By functional calculus, any function $g(H_\omega)$ is also a covariant operator. In particular, the Fermi projection $P_F = \{P_\omega=\chi(H_\omega< 0)\}_{\omega \in \Omega}$ or the flat band version $Q = \{Q_\omega={\rm sgn}(H_\omega)\}_{\omega \in \Omega}$ are covariant. 

\vspace{.1cm}

In the present context, each Hamiltonian $H_\omega$ is supposed to have the chiral symmetry \eqref{eq-ChiralHamiltonian} w.r.t. $J \otimes \bm 1$ with $J$ of the form \eqref{eq-Sblock}.  By functional calculus, it follows that also any odd function of $H_\omega$ has the chiral symmetry and thus, in particular, the flat band version $Q_\omega={\rm sgn}(H_\omega)$. As in \eqref{eq-Qblock}, it hence admits again a representation
\begin{equation}
\label{eq-Qblockcov}
Q_\omega
\;=\;
\begin{pmatrix}
0 & U_\omega^*
\\
U_\omega & 0
\end{pmatrix}
\;,
\end{equation}
with a unitary $U_\omega$ which satisfies again the covariance relation
\begin{equation}
\label{eq-covrelU}
V_{\bm a} U_\omega V_{\bm a}^\ast \;=\; U_{\action_{\bm a}\omega}
\;,
\end{equation}
where $V_{\bm a}$ is given by the same formula as above, albeit on the Hilbert space $\mathbb C^N \otimes \ell^2(\mathbb Z^d)$ with half-dimensional fiber. The family of Fermi unitary operators $U_F=\{U_\omega\}_{\omega\in\Omega}$ is thus precisely of the form required by Theorem~\ref{TheoIntro}. The condition \eqref{Co} holds either if the Fermi energy $E=0$ lies in a (almost sure) gap of the spectrum of $H_\omega$, or at least in a region of Anderson localization. This will be proved in Section~\ref{Sobolev}. In conclusion, if the localization condition holds, Theorem~\ref{TheoIntro} applies and allows to associate the odd Chern number and associated index to $U_F$ and thus to the ground state of the system. This provides the phase label discussed in the introduction.

\begin{example} 
An easy way to obtain an aperiodic chiral unitary system is to start from the periodic Hamiltonian of \eqref{LatticeH0} and define $H_\omega$ via \eqref{LatticeHamOmega} by setting $t_{\bm a}(\omega)=(1+\lambda_{\bm a}\, \omega^{\bm a}_{\bm 0})t_{\bm a}$, where $\lambda_{\bm a}>0$ are coupling constants and $\omega = \{\omega^{\bm a}_{\bm x}\}^{\bm a \in \mathcal R}_{\bm x \in \mathbb Z^d}$ is a collection of real numbers (such that $\omega^{\bm a}_{\bm x}=\omega^{-\bm a}_{\bm x}$) drawn independently according to some probability distribution with compact support, say uniformly from the interval $[-\frac{1}{2},\frac{1}{2}]$. Then $\omega \in \Omega=[-\frac{1}{2},\frac{1}{2}]^{\mathcal R \times \mathbb Z^{d}}$ and the latter is a compact Tychonov space which can be endowed with the probability measure $\mathbb P(d \omega)=\prod_{\bm x \in \mathbb Z^d} d \omega_x$ which is invariant and ergodic w.r.t. the natural shift action of $\mathbb Z^d$. To present something concrete, let us write out the disordered one-dimensional model analyzed numerically in Ref.~\cite{MSH} over $\ell^2(\mathbb Z^d,\mathbb C^2)$:
\begin{equation}
\label{Model2}
\begin{array}{c}
(H_\omega{\bm \psi})(x)\;=\;\frac{1}{2}(1+\lambda \omega_x)[(\sigma_1+\imath \sigma_2){\bm \psi}(x+1) \;+\; (\sigma_1-\imath \sigma_2){\bm \psi}(x-1)]\bigskip \\
 +\;(m+\lambda' \omega'_x) \sigma_2 {\bm \psi}(x),
 \end{array}
\end{equation}
with $\omega_x$ and $\omega'_x$ independent random variables uniformly distributed $[-\frac{1}{2},\frac{1}{2}]$. 
\hfill $\diamond$
\end{example}

\section{The algebras of covariant observables}
\label{Algebras}

Let us first introduce, following the work of Bellissard \cite{BellissardLNP1986jf},  the reduced crossed product algebras generating various classes of operators affiliated with the tight-binding models introduced in the previous section. Let $(\Omega,T,\mathbb{Z}^d,\mathbb P)$ be the C$^*$-dynamical system described in the introduction. Then the core algebra, denoted here by $\mathcal{A}_0$, is defined as the set $C_c(\Omega \times {\mathbb Z}^d)$ of compactly supported, continuous functions on $\Omega \times {\mathbb Z}^d$, furnished with the following algebraic operations:
\begin{align*}
(f+\lambda g)(\omega,{\bm x}) &\; =
\;f(\omega,{\bm x})\,+\,\lambda\,g(\omega,{\bm x})\;,  
\\
(fg)(\omega,{\bm x}) & \; =
\;\sum\limits_{{\bm y} \in {\mathbb Z}^d} e^{\imath {\bm y} \wedge {\bm x}}\,f(\omega, {\bm y})\,g(\action_{{\bm y}}^{-1}\omega,{\bm x}-{\bm y})\;, \\
f^\ast(\omega,{\bm x}) & 
\;=\;f(\action_{\bm x}^{-1}\omega,-{\bm x})^*\;. 
\end{align*} 
Covariant representations $\pi_\omega$ on $\ell^2(\mathbb Z^d)$ are defined by
$$
\mathcal A_0 \ni f 
\;\mapsto\;
\pi_\omega (f)=\sum_{{\bm y},{\bm x} \in {\mathbb Z}^d} e^{ \imath {\bm y} \wedge {\bm x}} f(\action_{\bm x}\omega, {\bm y}) |\bm x\rangle \langle \bm x|S_{\bm y}\;.
$$
The family $F=\{\pi_\omega (f)\}_{\omega \in \Omega}$ of such representations satisfies the covariance relation $V_a F_\omega V_a^*=F_{T_a \omega}$ with $V_a$ as defined in \eqref{eq-Vdef}. Inversely, given such a covariant family $F=\{ F_\omega \}_{\omega \in \Omega}$ of finite range operators, there is an associated $f\in\mathcal A_0$, defined by $f(\omega,{\bm x})= \langle{\bm 0}| F_\omega|-{\bm x}\rangle$, such that $F_\omega = \pi_\omega(f)$. As a general rule, the operators affiliated with the homogeneous tight-binding models can be obtained as representations of the matrix algebras $M_K(\mathbb C) \otimes \mathcal A_0$, and its closures discussed below, via the representations ${\rm id} \otimes \pi_\omega$. Here and throughout, $M_K(\mathbb C)$ denotes the algebra of $K\times K$ matrices with complex entries and, to simplify the notation, we will use $\pi_\omega$ instead of ${\rm id} \otimes \pi_\omega$. For example, the family of covariant Hamiltonians $H=\{H_\omega\}_{\omega\in\Omega}$ defined in \eqref{LatticeHamOmega} can be identified with an element $h\in M_{2N}(\mathbb C) \otimes \mathcal A_0$ via $h(\omega,{\bm x})=t_{\bm x} (\omega)$. The spectrum of $h$ and the almost sure spectrum of $H$ coincide. The core algebra can be closed w.r.t. the C$^\ast$-norm
$$
\|f\|\; =\; \sup_{\omega \in \Omega} \|\pi_\omega f \|
\;.
$$
and the result is a C$^*$-algebra $\mathcal A$ which is nothing but the reduced twisted crossed product algebra \cite{Ped} associated to $(\Omega,T,\mathbb{Z}^d,\mathbb P)$. If $h$ displays a spectral gap at zero energy, then $p_F = \chi(h < 0)$ as well as $q = {\rm sgn}(h)$ are well defined via the continuous functional calculus in $M_{2N}(\mathbb C) \otimes \mathcal A$ (by using a harmless smoothing of $\chi$ and ${\rm sgn}$). These elements generate through $\pi_\omega$ the Fermi projection $P_F$ and the flat band version $Q$ of $H$. Furthermore, due to the chiral symmetry, $q$ necessarily takes the form
$$
q 
\;=\; 
\begin{pmatrix} 0 & u_F^* \\ u_F & 0 \end{pmatrix}
\;,
$$
which enables us to identify the Fermi unitary element $u_F \in M_{N}(\mathbb C) \otimes \mathcal A$, which generates $U_F$ through representations $\pi_\omega$. If there is no spectral gap at the Fermi level, the smoothing trick no longer works and $p_F$ and $u_F$ are no longer in the C$^\ast$-algebra which is only stable under the continuous functional calculus. In this case, one has to consider the von Neumann closure $L^\infty(\mathcal A,\mathbb P)$ of $\mathcal A_0$, under the norm
$$
\|f\|_\infty\; =\; 
\mathbb P\!-\!\operatorname*{\mathrm{esssup}}\limits_{\omega \in \Omega} \|\pi_\omega (f)\|
\;.
$$
This algebra is stable under the Borel functional calculus and therefore $M_{2N}(\mathbb C) \otimes L^\infty(\mathcal A,\mathbb P)$ contains ${\rm sgn}(h)$, regardless of the existence of a spectral gap at the Fermi energy. Using the chiral symmetry, we can identify again the Fermi unitary element, but this time $u_F$ belongs to $M_N(\mathbb C) \otimes L^\infty(\mathcal A,\mathbb P)$. Further essential properties of $u_F$ are discussed below.

\vspace{.1cm}

The algebra $\mathcal A_0$ and its completions become non-commutative manifolds when equipped with non-commutative differential calculus tools, namely integration
$$
\mathcal{T}(f)\;=\;\int_\Omega \mathbb P (d\omega) \, f(\omega,{\bm 0})
\;, 
$$
and derivations
$$
(\partial_j f)(\omega,{\bm x}) \;=\; - \imath\, x_j\, f(\omega,{\bm x})\;,
\qquad j=1,\ldots,d\;,
$$
where $\bm x=(x_1,\ldots,x_d)$. The linear functional $\mathcal T$ defines a continuous, faithful and normalized trace over $\mathcal A$ and a tracial state over $L^\infty(\mathcal A,\mathbb P)$. One can consider the Banach spaces $L^p(\mathcal A,\mathcal T)$  defined as the closure of $\mathcal A_0$ under the norm 
\begin{equation}
\label{LPNorm}
\|f\|_p 
\;=\; 
\mathcal T(|f|^p)^\frac{1}{p}
\;,
\end{equation}
in which case $L^\infty(\mathcal A,\mathbb P)$ can be equivalently characterized as $L^\infty(\mathcal A,\mathcal T)$ \cite{Con3}. In the following we will adopt the latter notation. A useful tool at several instances is the non-commutative version of H\"older's inequality
\begin{equation}
\label{Holder}
 \|f_1 \cdots f_k\|_p 
 \;\leq \;
 \|f_1\|_{p_1} \cdots \| f_k \|_{p_k}
\;, 
\qquad \frac{1}{p_1} + \ldots + \frac{1}{p_k} = \frac{1}{p} 
\;.
\end{equation}
 
Next, let us define the smooth sub-algebra $\mathscr A$ inside the C$^\ast$-algebra $\mathcal A$ (in the sense of \cite{Ren}). We will denote by $C^n(\mathcal A)$ the algebra of $n$-times differentiable elements, {\it i.e.} those $a\in\mathcal A$ for which $\bm \partial^{\bm \alpha} a \in \mathcal A$, for all multi-indices $\bm \alpha = (\alpha_1,\ldots,\alpha_d)$ with $|\bm \alpha|=\alpha_1 + \ldots + \alpha_d \leq n$. Then $\mathscr A \subset \mathcal A$ is given by the space of infinitely differentiable elements
$$
\mathscr A
\;=\;
C^\infty(\mathcal A)
\;=\;
\bigcap_{n\geq 0} C^n(\mathcal A)
\;,
$$
equipped with the convex topology induced by the seminorms
$$
\|f \|_{\bm \alpha} \;=\; \|\bm \partial^{\bm \alpha} f \|\;, 
\qquad 
\bm \partial^{\bm \alpha} = \partial_1^{\alpha_1} \cdots \partial_d^{\alpha_d}
\;,\;\;
\bm \alpha \;=\; 
(\alpha_1, \ldots \alpha_d)
\;,
$$ 
where the norm appearing on the right is the C$^\ast$-norm of $\mathcal A$. We recall \cite{Con2} that $\mathscr A$ is a dense Fr\'{e}chet sub-algebra of $\mathcal A$ which is stable under holomorphic calculus. An important consequence of this is that the $K$-theories of the two algebras coincide \cite{Con1}. If a spectral gap is present at the Fermi level, then standard techniques can be used to show that the Fermi unitary element $u_F$ belongs to $M_N(\mathbb C)\otimes \mathscr A$. Lastly, $\mathscr A$ can be characterized as the sub-algebra of elements with rapid decay in the second variable, more precisely:

\begin{proposition} If $f \in \mathscr A \subset \mathcal A$, then for any multi-index $\bm \alpha$:
\begin{equation}
\label{Decay1}
\bm x^{\bm \alpha}\sup_{\omega \in \Omega} 
|f(\omega,\bm x)| 
\;\leq\; 
\|\bm \partial^{\bm \alpha} f\| 
\;<\; \infty\;, 
\qquad 
\bm x^{\bm \alpha} = x_1^{\alpha_1} \cdots \bm x_d^{\alpha_d}
\;.
\end{equation}
\end{proposition}

\proof One has
$$
|\bm x^{\bm \alpha}f(\omega, \bm x)| 
\;=\; 
| \langle \bm 0 |\pi_\omega(\bm \partial^{\bm \alpha} f) | \bm x \rangle |\leq  \sup_{\omega \in \Omega} \|\pi_\omega(\bm \partial^{\bm \alpha} f)\| 
\;=\; 
\|\bm \partial^{\bm \alpha} f\| 
\;<\; \infty
\;,
$$
which concludes the proof. \qed

\vspace{0.2cm}

As already mentioned, in the strong disorder regime when there is merely a mobility gap at the Fermi level, $u_F$ is not even an element of $M_N(\mathbb C)\otimes \mathcal A$ and, instead, it belongs to the weak von Neumann closure $L^\infty(\mathcal A,\mathcal T)$. However, the topology of this algebra is too strong and $u_F$ does not vary continuously w.r.t.  norm $\| \cdot \|_\infty$, even when the models are deformed continuously. Hence, in order to analyze the strong disorder regime, we need to determine the appropriate space of elements. In \cite{BELLISSARD:1994xj,ProdanJPA2013hg}, this space was identified as a certain non-commutative Sobolev space. These Banach spaces, denoted here by $\mathcal W_{p,r}(\mathcal A,\mathcal T)$, are defined as the closure of $\mathcal A_0$ under the norms
\begin{equation}
\label{SobolevNorm}
\|f\|_{p,r}
\;=\; 
\sum_{|\bm \alpha|\leq r} {\mathcal T}\left ( |\bm \partial^{\bm \alpha} f|^p\right )^{\frac{1}{p}}
\;=\; 
\sum_{|\bm \alpha|\leq r} \|\bm \partial^{\bm \alpha} f\|_p
\;.
\end{equation}
The use of $\mathcal W_{p,r}(\mathcal A,\mathcal T)$ in \cite{ProdanJPA2013hg} depended crucially on the ability to compute a certain Dixmier trace in order to prove the summability property discussed in the next section. To avoid this highly technical step, which is further complicated by the odd dimensionality of the space, we proceed here in a different way by defining a new class of Banach spaces given by the completion of $\mathcal A_0$ under the norms
$$
\|f\|'_{p,r} 
\;=\; 
\sum_{\bm x \in \mathbb Z^d} (1+|\bm x|)^r \left [ \int_\Omega \mathbb P(d \omega) |f(\omega,\bm x)|^p \right ]^\frac{1}{p}
\;, 
\qquad 
p \in [1,\infty)\;, \;\; r\in [0,\infty)
\;. 
$$
Due to similarities which will be highlighted below, we will call them Sobolev spaces as well and will use the notation $\mathcal W'_{p,r}(\mathcal A,\mathbb P)$. Some of their important properties are listed below.

\begin{proposition}\label{SobolevProp} The new Sobolev spaces satisfy the following relations:

\begin{enumerate}[\rm (i)]

\item $\mathcal W'_{p,r}(\mathcal A,\mathbb P)$ is invariant to the $\ast$-operation.

\item $\mathcal W'_{p,r}(\mathcal A,\mathbb P) \subset \mathcal W'_{p',r'}(\mathcal A,\mathbb P)$, whenever $p \leq p'$ and $r \leq r'$.

\item $\mathcal W_{p,r}(\mathcal A,\mathcal T) \subset \mathcal W'_{p,r}(\mathcal A,\mathbb P)$, for $p$ an even integer.

\end{enumerate}

\end{proposition}

\proof (i) Since the norms $\| \cdot \|'_{p,r}$ are invariant to the transformations $f(\omega,\bm x) \rightarrow f(T_{\bm y}\omega,-\bm x)$ and to the complex conjugation of $f(\omega,\bm x)$, the equality $\|f^\ast \|_{p,r} = \|f\|_{p,r}$ holds. (ii) Clearly, 
\begin{equation}
\label{Comp1}
\|f\|'_{p,r} 
\;\leq\; 
\|f\|'_{p,r'}
\;, 
\qquad {\rm for} \ r \leq r'
\;,
\end{equation}
and H\"older's inequality gives $\|f\|'_{p,r} \leq \|f\|'_{p',r}$ whenever $p < p'$. (iii) We will show
\begin{equation}
\label{NormInequality}
\|f\|_{p,r} 
\;\leq\; 
\mathcal N_r \,\|f\|'_{p,r}
\;,
\end{equation}
with a constant $\mathcal N_r$ specified below. We start from the l.h.s. and evaluate first the $L^p$-norm $\|\bm \partial^{\bm \alpha} f\|_p$ from \eqref{LPNorm}. The difficult part here is to compute the absolute value, but for $p$ even integer we can proceed as
$$
\|\bm \partial^{\bm \alpha} f\|_p^p 
\;=\;
\mathcal T\Big (\Big ( (\bm \partial^{\bm \alpha} f^*) (\bm \partial^{\bm \alpha} f ) \Big )^\frac{p}{2} \Big )
\;.
$$
By writing out the products explicitly,
\begin{align*} 
\|\bm \partial^{\bm \alpha} f\|_p^p  
\;\leq\; 
\int_\Omega \mathbb P(d\omega) \sum_{\bm x_2,\ldots,\bm x_p \in \mathbb Z^d}    \prod_{i=1}^p |(\bm x_{i+1}-\bm x_i)^{\bm \alpha} 
f^\#(T_{\bm x_i} \omega,\bm x_{i+1} - \bm x_i) |
\;,
\end{align*}
where $\bm x_1$ and $\bm x_{p+1}$ are pinned to the origin and  $\#$ indicates that some of the $f'$s are adjoint. Since there are only positive terms, the sums and the integral can be exchanged. Then the changes of variables $\bm y_i = \bm x_{i+1} - \bm x_i$, $i=1,\ldots,p$, together with H\"older's inequality give
\begin{align*} 
\|\bm \partial^{\bm \alpha} f\|_p^p 
& \;\leq \;
\sum_{\bm y_1+ \ldots + \bm y_p =0}  \ \prod_{i=1}^p |\bm y_i^{\bm \alpha}| \left [ \int_\Omega \mathbb P(d\omega) |f^\#(T_{\bm x_i} \omega,\bm y_i) |^p \right ]^\frac{1}{p} 
\\
& 
\;\leq\; 
\left ( \sum_{\bm y \in \mathbb Z^d} (1+|\bm y|)^{|\bm \alpha|} \left [ \int_\Omega \mathbb P(d\omega) |f(\omega,\bm y) |^p \right ]^\frac{1}{p} \right )^p.
\end{align*}
or $\|\bm \partial^{\bm \alpha} f\|_p  \leq \|f\|'_{p,|\bm \alpha|}.$ Then
$$
\|f\|_{p,r} 
\;=\;
\sum_{|\bm \alpha|\leq r} \|\bm \partial^{\bm \alpha}f\|_p 
\;\leq\; 
\sum_{|\bm \alpha|\leq r} \|f\|'_{p,|\bm \alpha|} 
\;\leq \;
\mathcal N_r \|f\|'_{p,r}
\;,
$$
where $\mathcal N_r$ is the cardinality of the set $\{\bm \alpha \in \mathbb N^d\,:\,  |\bm \alpha| \leq r\}$. \qed

\vspace{.2cm} 

\section{Implementing Connes' program}
\label{ACProgram}

This section introduces a natural field of finitely summable odd Fredholm modules over the C$^\ast$-algebra $\mathcal A$ of covariant observables (more precisely $\mathscr A$) and derives a local formula for its Chern character.  Our calculations do not make appeal to the general Connes-Moscovici local formula \cite{CM} and instead rely on the geometric identity of Lemma~\ref{KeyIdentity}. After a preprint of this work was made available, the Connes-Moscovici local formula for the same Chern character was evaluated \cite{Bou} by employing tools from \cite{CGX}, and it was shown that this route leads to the same result. From the physics point of view, the main output of this section is an index theorem for the odd Chern number defined in \eqref{OddChernNr1bis} which ensures the quantization and invariance of this topological invariant, but only under a spectral gap assumption. This assumption will be relaxed in next section.

\subsection{A finitely summable odd Fredholm module}

 The standard terminology from \cite{Con2} will be adopted throughout. As in the introduction, let $\sigma_1, \ldots, \sigma_d$ be the generators of an irreducible representation of the odd complex Clifford algebra $Cl_{d}$ on representation space $\mathrm{Cliff}(d)$ of dimension $d'=2^{\frac{d-1}{2}}$, namely 
 $$
 \sigma_i \sigma_j + \sigma_j \sigma_i = 2 \delta_{i,j}, \quad \sigma_j^* = \sigma_j, \quad i,j=1,\ldots d \;.
 $$ 
 On the augmented Hilbert space ${\mathcal H}=\mathrm{Cliff}(d) \otimes \ell^2({\mathbb Z}^d)$, the shifted Dirac operator is introduced by
$$
D_{\bm x_0}
\;=\;
\sum_{j=1}^d \sigma_j \otimes (X_j +x_{0,j})
\;,
$$
where $\bm x_0=(x_{0,1},\ldots,x_{0,d})\in \mathbb R^d$.  If $\bm x_0 \in \mathbb Z^d$, the Dirac operator has zero modes localized at $\bm x = - \bm x_0$ which are eliminated by a  finite rank modification upon setting, for example, $(D_{\bm x_0}\bm \psi)(-\bm x_0) = \bm \psi(-\bm x_0) $. This allows us to introduce the Dirac phase
$$
F_{{\bm x}_0}
\;=\;
\frac{D_{{\bm x}_0}}{|D_{{\bm x}_0}|}
\;.
$$
Furthermore, the elements of $\mathcal A$ can be represented on $\mathcal H$ by ${\rm id} \otimes \pi_\omega$. Consistent with our previous conventions, $\pi_\omega$ will be used instead of ${\rm id} \otimes \pi_\omega$.

\begin{proposition} 
\label{Summability}
The triples $({\mathcal H},F_{{\bm x}_0},\pi_\omega)$, with ${\bm x}_0$ in the unit cube $\mathcal C^d = [0,1]^d$ and ${\omega \in \Omega}$, define a field of $(d+1)$-summable odd Fredholm modules over the smooth sub-algebra $\mathscr A$.
\end{proposition}

\proof We only need to verify the $(d+1)$-summability condition
\begin{equation}
\label{Summability1}
{\rm Tr}\left (\big |[F_{\bm x_0}, \pi_\omega(f)]\big |^{d+1} \right ) \;<\; \infty\;, 
\qquad 
\forall\;\; f \in \mathscr A
\;.
\end{equation}
Due to the Minkovski inequality for the Schatten norms and the decomposition $f=\frac{1}{2}(f+f^*)+\frac{1}{2}(f-f^*)$, it is sufficient to consider the case $f=f^*$. Let $|\bm x,n \rangle$, $\bm x \in \mathbb Z^d$ and $n = 1,\ldots,d'$, denote the canonical basis for $\mathcal H$. Also $\langle \bm x | \cdot |\bm x' \rangle$ is understood as the matrix with the entries $\langle \bm x,n | \cdot |\bm x',m \rangle$ and $| \cdot |$ will denote its matrix norm. We first show that
\begin{equation}
\label{S1}
\sum_{\bm x' \in \mathbb Z^d}\big | \langle \bm x' | [F_{\bm x_0}, \pi_\omega(f)]^k| \bm x \rangle \big | 
\;\leq\; 
A |\bm x + \bm x_0|^{-k}
\;,
\end{equation}
for any positive integer k. Throughout, we will specify all non-interesting, but finite constants by $A$. It will be convenient to introduce the notation $\widehat{\bm v} = \bm v /|\bm v|$ for the non-zero vectors from $\mathbb R^d$, and it will be helpful to have the commutators written explicitly 
$$
\langle \bm x|   [F_{\bm x_0}, \pi_\omega(f)]| \bm y \rangle  
\;=\;  
e^{\imath \bm x \wedge \bm y } f(T_{\bm x} \omega, \bm x-\bm y) \bm \; \bm \sigma \cdot  \big (\widehat{\bm x + \bm x_0} - \widehat{\bm y + \bm x_0} \big )
\,.
$$
Now let us denote the l.h.s. of \eqref{S1} by $Y$. Then, after a lattice shift by ${\bm x}$ and with the help of the above formula,
\begin{align*}
 Y \;\leq\; 
 \sum_{\bm x_1,\ldots,\bm x_{k+1} \in \mathbb Z^d}\, \delta_{\bm x_1, \bm 0} \prod_{i=1}^k \; |f(\omega_i, \bm x_{i+1}-\bm x_{i})| & \big |\widehat{\bm x_{i+1} +\bm x + \bm x_0} - \widehat{\bm x_{i} + \bm x + \bm x_0} \big |
\;,
\end{align*}
where $\omega_i = T_{\bm x_{i+1}-\bm x} \omega$. Given the asymptotic behavior for $|\bm x| \rightarrow \infty$,
\begin{equation}\label{Asy}
\widehat{\bm x + \bm y}  - \widehat{\bm x + \bm y'} 
\;\sim \;
|\bm x|^{-1}  \Big ( \bm y - \bm y' +\big ( \widehat{\bm x}\cdot (\bm y - \bm y') \big )  \widehat{ \bm x} \Big )
\;,
\end{equation}
the supremum 
$$
S(\bm y,\bm y')=\sup_{\bm x \in \mathbb R^d} \;|\bm x|\, \big | \widehat{\bm x + \bm y}- \widehat{\bm x + \bm y'} \big |
$$
is finite. It clearly posses the homogeneity property $ S(s \bm y, s \bm y') = s\,S(\bm y, \bm y')$, hence, by taking $s = (|\bm y|+|\bm y'|)^{-1}$, we obtain
$$
S(\bm y,\bm y') 
\;\leq \;
(|\bm y|+|\bm y'|) \sup_{|\bm x|+|\bm x'| = 1} \;S(\bm x,\bm x')
\;.
$$
Consequently
\begin{align*}
Y 
\;\leq \;
 A |\bm x + \bm x_0|^{-k} \sum_{\bm x_1,\ldots,\bm x_{k+1} \in \mathbb Z^d}\, \delta_{\bm x_1, \bm 0}\prod_{i=1}^k \;(| \bm x_i|+|\bm x_{i+1}|)\; |f(\omega_i, \bm x_{i+1}-\bm x_{i})|
\;. \nonumber
\end{align*} 
We now make the change of variables $\bm y_i = \bm x_{i+1}-\bm x_i $ for $i=1,\ldots,k$, and observe that, since $\bm x_1=\bm 0$,
$$
\bm x_{i+1} 
\;=\;
\bm y_i + \ldots + \bm y_1
\;\; \Longrightarrow \;\;
|\bm x_{i+1}| \leq \prod_{j=1}^k (1+|\bm y_j|)
\;.
$$
This gives
\begin{align}\label{KPower}
Y 
\; \leq \;
 A |\bm x + \bm x_0|^{-k} 
\,\sum_{\bm y_1,\ldots,\bm y_k \in \mathbb Z^d}  
\prod_{i=1}^k \;(1+|\bm y_i|)^k\; |f(\omega_i, \bm y_i)|
\;,
\end{align} 
and furthermore
\begin{align*}
Y 
\; \leq \;
 A |\bm x + \bm x_0|^{-k} \Big (\sum_{\bm y \in \mathbb Z^d}(1+|\bm y|)^k\; \sup_{\omega \in \Omega} |f(\omega, \bm y)| \Big )^k.
\end{align*} 
The sum is finite due to \eqref{Decay1} and this concludes the proof of \eqref{S1}. Now we take $k=d+1$, which is an even number. As $f$ is self-adjoint, 
$$
\big |[F_{\bm x_0}, \pi_\omega(f)] \big |^{d+1}
\;=\; 
(\imath [F_{\bm x_0}, \pi_\omega(f)])^{d+1} 
\;,
$$
and, from Eq.~\eqref{S1},
$$
{\rm Tr}\left (\big |[F_{\bm x_0}, \pi_\omega(f)]\big |^{d+1} \right ) 
\;\leq \;
A\, \sum_{\bm x \in \mathbb Z^d} |\bm x +\bm x_0|^{-d-1} 
\;<\; \infty
\;,
$$
and $(d+1)$-summability follows. \qed

\subsection{The Chern character and its paring with the $K_1$ group}
\label{sec-Chern}

To each of the odd $(d+1)$-summable odd Fredholm modules $(\mathcal H, F_{\bm x_0},\pi_\omega)$ over $\mathscr A$, one can define in a standard manner Connes' cyclic $(d+1)$-cocycle \cite{Con2}, inside the cyclic cohomology of $\mathscr A$ \cite{Con1},
\begin{equation}
\label{ChernCharacter1}
\tau_{d,\omega,\bm x_0}(f_0, \ldots,f_d)
\; = \;
\lambda_d\, \mathrm{Tr}
\left(
F_{\bm x_0}[F_{\bm x_0},\pi_\omega (f_0)] \cdots [F_{\bm x_0},\pi_\omega(f_d)]
\right)
\;,
\end{equation}
where $\lambda_d\,=\,\imath^{d+1}/2^{d+1}$ is chosen such that \eqref{ChernIndex} below holds. The cohomology class $[\tau_{d,\omega,\bm x_0}]$ of $\tau_{d,\omega,\bm x_0}$ defines the Chern character of the Fredholm module $({\mathcal H},F_{{\bm x}_0},\pi_\omega)$. We define the Chern character of the entire family of Fredholm modules as the class of the following cyclic-cocycle
$$
\tau_d(f_0,\ldots,f_d) 
\;=\; \int_{\mathcal C^d} d \bm x_0 \int_\Omega \mathbb P(d\omega)  \; \tau_{d,\omega,\bm x_0}(f_0, \ldots,f_d)
\;.
$$

Recall now that the topological $K_1$ group of the C$^\ast$-algebra $\mathcal A$ \cite{RLL} is given by the quotient $K_1(\mathcal A)= U_\infty(\mathcal A)/ U_\infty(\mathcal A)_0$ where $U_\infty(\mathcal A)$ is the inductive limit of the groups of unitary elements from $M_N(\mathbb C) \otimes \mathcal A$ and $U_\infty(\mathcal A)_0$ is its connected component of the unity. The topological $K_1$ group can be defined in the same manner for the Fr\'{e}chet sub-algebra $\mathscr A$, too, and, since $\mathscr A$ is dense and stable under the holomorphic calculus, $K_1(\mathscr A) = K_1(\mathcal A)$ \cite{Con1}. The following proposition summarizes some of the fundamental statements in non-commutative geometry \cite{Con2}.

\begin{proposition}[\cite{Con2}] 
\label{ChernCharacter}
Let ${\rm Tr} \, \# \, \tau_d$ be the standard extension of the cocycle over $M_{\infty}(\mathbb C) \otimes \mathscr A$. Then:

\begin{enumerate}[\rm (i)]

\item The map
\begin{equation}
\label{Cocycle1}
u\in U_\infty(\mathscr A) \;\mapsto\; ({\rm Tr} \, \# \, \tau_d) (u^\ast-1,u-1,\ldots,u^\ast-1,u-1)
\;,
\end{equation}
is constant on the equivalence class $[u]$ of $u$ in $K_1 (\mathscr A)$. 

\item This constant value remains unchanged if $\tau_d$ is replaced by any other representative from its cohomology class. Hence, {\rm \eqref{Cocycle1}} defines a natural pairing  $\langle [\tau_d], [u] \rangle$ between  $K_1 (\mathscr A )$ $(=K_1(\mathcal A))$ and the Chern character.

\item The pairing is integral and given by the index of a Fredholm operator. Specifically, if $u$ is a unitary element from  $M_N(\mathbb C) \otimes \mathscr A$, then
\begin{equation}
\label{ChernIndex}
\langle [\tau_d],[u] \rangle 
\;= \;
\mathrm{Ind} \left( E_{\bm x_0} \pi_\omega (u) E_{\bm x_0}\right) \,\in\, \mathbb{Z},
\end{equation}
where $E_{\bm x_0}$ is the projection $\bm 1 \otimes \frac{1}{2}(\bm 1+F_{\bm x_0})$ on $\mathbb C^N \otimes \mathcal H$. The Fredholm index at the r.h.s. of \eqref{ChernIndex} is independent of $\omega$ and $\bm x_0$.
\end{enumerate}
\end{proposition}

\vspace{.2cm}

The main result of this section is a local formula for the Chern character.

\begin{theorem}
\label{LocalFormula} 
For $f_0,\ldots,f_d\in \mathscr A$,
\begin{equation}
\tau_d(f_0,\ldots,f_d)
\;  =\; 
\Lambda_d \sum_{\rho \in \mathcal S_d}(-1)^\rho\; {\mathcal T} \Big ( f_0\prod_{i=1}^d \partial_{\rho_i}f_i\Big )
\;,
\qquad
\Lambda_d
\;=\;
\frac{\imath\,(\imath \pi)^{\frac{d-1}{2}}}{d!!}
\;.
\label{eq-theolocal}
\end{equation}
\end{theorem}

\proof Let us write the cocycle $\tau_d=\tau_d(f_0,\ldots,f_d)$ explicitly:
\begin{equation}
\label{Cocycle}
\tau_d
\; = \;
\lambda_d\, \int_{\mathcal C^d} d \bm x_0 \int_\Omega \mathbb P(d\omega)  \;\mathrm{Tr}
\big (
F_{\bm x_0}[F_{\bm x_0},\pi_\omega (f_0)] \cdots [F_{\bm x_0},\pi_\omega(f_d)]
\big )
\;.
\end{equation}
Due to the summability property proved in Proposition~\ref{Summability}, the sums and integrals may be exchanged in the calculations below. As for the notation, $\mathrm{tr}_{\sigma}$ will denote the trace over the fiber $\mathrm{Cliff}(d)$. Now, let us write the trace explicitly and use the lattice translations and the covariance property to move all the fibers to the origin
\begin{align*}
\tau_d
& \;=\;
\lambda_d \int_{{\mathcal C}^d} d{\bm x}_0 \int_\Omega \mathbb P (d\omega) \; 
\sum\limits_{{\bm x} \in {\mathbb Z}^d} \mathrm{tr}_{\sigma} \Big (\langle \bm 0 |F_{{\bm x}_0+{\bm x}} \prod_{i=0}^d [F_{\bm x_0 + \bm x},\pi_{T_{\bm x}\omega}( f_i)]  | \bm 0 \rangle \Big)
\\
& \;=\;
\lambda_d \int_{\mathbb{R}^d} d{\bm x} \int_\Omega \mathbb P (d\omega) \; 
\mathrm{tr}_{\sigma}
\Big ( \langle \bm 0 | F_{{\bm x}} \prod_{i=0}^d [F_{\bm x},\pi_{\omega}( f_i)]  | \bm 0 \rangle \Big )
\;,
\end{align*}
where in the second equality the invariance of ${\mathbb P}$ was used. Breaking the commutator $[F_{\bm x},\pi_{\omega}( f_0)]$ and using $F_{{\bm x}}[F_{{\bm x}},\pi_\omega( f)]=-[F_{{\bm x}},\pi_\omega (f)]F_{{\bm x}}$ as well as $F_{{\bm x}}^2={\bf 1}$ then leads to
$$
\tau_d
\;=\;
2 \lambda_d \int_{\mathbb{R}^d} d{\bm x} \int_\Omega \mathbb P (d\omega) \; 
\mathrm{tr}_{\sigma} \Big ( \langle \bm 0 | \pi_\omega (f_0) \prod_{i=1}^d [F_{{\bm x}},\pi_\omega (f_i) ] | \bm 0 \rangle  \Big )
\;.
$$
Next let us insert partitions of unity, using the projections $\chi_{\bm x}=|\bm x \rangle \langle \bm x |$,
$$
\tau_d
\,=\,
2\lambda_d \int_{\mathbb{R}^d} d{\bm x} \int_\Omega \mathbb P (d\omega) \!
\sum\limits_{{\bm x}_1,\ldots,\bm x_d \in {\mathbb Z}^d} \!\mathrm{tr}_{\sigma} \!
\Big (\langle \bm 0 | \pi_\omega(f_0) \prod_{i=1}^d\chi_{{\bm x}_i}[F_{{\bm x}},\pi_\omega (f_i)]\chi_{{\bm x}_{i+1}} | \bm 0 \rangle \Big )
\,,
$$
where ${\bm x}_{d+1}$ is fixed at the origin. Interchanging the sum over ${\bm x}_i$ with the integrals allows to continue to
\begin{align*}
\tau_d
& \;=\; 2 \lambda_d \sum\limits_{{\bm x}_1,\ldots,\bm x_d \in {\mathbb Z}^d} \int_{\mathbb{R}^d} d{\bm x} \;
\mathrm{tr}_\sigma \Big ( \prod_{i=1}^d {\bm \sigma} \cdot \big (\widehat{{\bm x}_i + {\bm x}} - \widehat{{\bm x}_{i+1}+{\bm x}} \big )  \Big )
\; 
\\
& \;\;\;\;\;\;\;\;\;
\times \int_\Omega \mathbb P (d\omega) \, \langle \bm 0 | \pi_\omega( f_0)\prod_{i=1}^d\chi_{{\bm x}_i} \, \pi_\omega(f_i) \, \chi_{{\bm x}_{i+1}}  | \bm 0 \rangle
\;.
\end{align*}
Now the integral in the first line can be evaluated with the identity from Lemma~\ref{KeyIdentity}. Expressing the sums over ${\bm x}_i$ with the position operators $X_i$ and connecting the constants $\lambda_d$ and $\Lambda_d$ then leads to
$$
\tau_d
\;=\;  
\imath^{d}\, \Lambda_d \,\int_\Omega \mathbb P (d\omega) \sum_{\rho\in \mathcal S_d} (-1)^\rho\; 
\langle \bm 0 | \pi_\omega(f_0) \prod_{i=1}^d X_{\rho_i}\pi_\omega (f_i) | \bm 0 \rangle
\;.
$$
Due to the anti-symmetrizing factor $(-1)^\rho$, one can actually form commutators
$$
\tau_d
\;=\;  
\imath^{d}\,
\Lambda_d \,
 \sum_{\rho\in \mathcal S_d} (-1)^\rho\; 
\int_\Omega \mathbb P (d\omega) \; \langle \bm 0 | \pi_\omega(f_0) \prod_{i=1}^d [X_{\rho_i},\pi_\omega(f_i)]  | \bm 0 \rangle
\;.
$$
This expression can finally be rewritten with the non-commutative analysis tools to complete the proof of the identity \eqref{eq-theolocal}.
\qed 

\begin{lemma}[Geometric identity] 
\label{KeyIdentity}
Let ${\bm x}_1, \ldots, {\bm x}_{d+1}\in \mathbb{R}^d$ with ${\bm x}_{d+1}={\bm 0}$ fixed at the origin. Then:
\begin{equation}
\label{KIdentity}
\int\limits_{\mathbb{R}^d} d{\bm x} 
\;
\mathrm{tr}_\sigma \Big (\prod_{i=1}^d {\bm \sigma} \cdot (\widehat{{\bm x}_i + {\bm x}} - \widehat{{\bm x}_{i+1}+{\bm x}} ) \Big)
\;=\;
-\frac{2^d (\imath\pi)^\frac{d-1}{2}}{d !!} \sum_{\rho \in \mathcal S_d} (-1)^\rho \prod_{i=1}^d x_{i,\rho_i} 
\;,
\end{equation}
where $x_{i,j}$ denotes the $j$-th component of ${\bm x}_i=(x_{i,1},\ldots,x_{i,d})$.
\end{lemma}

\noindent {\bf Remark.} A similar identity for even dimensions and even Clifford algebras has been found in \cite{ProdanJPA2013hg}.
\hfill $\diamond$

\proof Let us first list several useful identities for the $\sigma$ matrices:
\begin{enumerate}[(i)]
\item $\sigma_1 \sigma_2 \cdots \sigma_d = \pm \, \imath^\frac{d-1}{2} {\bf 1}$ (we choose the representation such that ``+" occurs);
\item $\mathrm{tr}_\sigma(\sigma_{\rho_1} \cdots \sigma_{\rho_q})=0$ if $q$ odd and $q<d$;
\item $\mathrm{tr}_\sigma (\sigma_{\rho_1} \cdots \sigma_{\rho_d} )=0$ unless $\rho$ is a permutation of $1,2,\ldots,d$;
\item $\mathrm{tr}_\sigma (\sigma_{\rho_1} \cdots \sigma_{\rho_d})= (2\imath)^\frac{d-1}{2} (-1)^\rho$ if $\rho$ is such a permutation.
\end{enumerate}
All these identities follow from the defining relations of the Clifford algebra. Notice that the product in the first identity commutes with all $\sigma$'s hence it must be proportional to the identity matrix.  Also, the square of the product equals $(-1)^\frac{d-1}{2}{\bf 1}$, which enables one to establish the constant in the first identity. The second identity follows from the fact that a conjugation with a $\sigma$ which is absent in that product of $\sigma$'s (here is where the condition $q<d$ enters) changes the sign of the trace but in the same time the trace is invariant to conjugations. If $\rho$ is not a permutation in the third identity, then pairs of identical $\sigma$'s can be erased from the product  and the identity then follows from the second identity. The fourth identity is evident.

\vspace{.1cm}

Using these identities, one can next establish 
\begin{equation}
\label{Id1}
\mathrm{tr}_\sigma \Big (\prod_{i=1}^d ({\bm \sigma} \cdot {\bm y}_i) \Big)
\;=\;
(2\imath)^\frac{d-1}{2}  d!  \ \mathrm{Vol}[{\bm 0},{\bm y}_1,\ldots,{\bm y}_d]
\;,
\end{equation}
for any set of points ${\bm y}_1,\ldots,{\bm y}_d\in \mathbb{R}^d$ and with $[{\bm y}_0,{\bm y}_1,\ldots,{\bm y}_d]$ denoting the corresponding simplex, and $\mathrm{Vol}[\ldots]$ the oriented volume of the simplex. Indeed, expending and taking into account (iii), one has
$$
\mathrm{tr}_\sigma \Big (\prod_{i=1}^d ({\bm \sigma} \cdot {\bm y}_i) \Big )
\;=\;
\sum_{\rho \in \mathcal S_d} y_{1,\rho_1} \cdots y_{d,\rho_d} \; \mathrm{tr}_\sigma 
\left(\sigma_{\rho_1} \cdots \sigma_{\rho_d} \right)
\;,
$$
and from (iv)
$$
\mathrm{tr}_\sigma \left(\prod_{i=1}^d ({\bm \sigma} \cdot {\bm y}_i) \right)
\;=\; 
(2 \imath)^\frac{d-1}{2} \;\mathrm{Det}[{\bm y}_1,\ldots,{\bm y}_d]
\;,
$$
where inside the determinant is the $d\times d$-matrix of columns ${\bm y}_1, \ldots, {\bm y}_d$. The volume of a simplex can be computed with the formula:
$$
\mathrm{Vol}[{\bm y}_0,{\bm y}_1,\ldots,{\bm y}_d]
\; =\;
 \frac{1}{d!}\;\mathrm{Det}\left [
\begin{array}{cccc}
{\bm y}_0 & {\bm y}_1   & \ldots & {\bm y}_d \\
1  &  1 &  \ldots & 1 
\end{array}
\right ]
\;,
$$
hence Eq.~\ref{Id1} follows by setting $\bm y_0 = \bm 0$ above.

\vspace{.1cm}

For the computation of the l.h.s. of \eqref{KIdentity} let us set
$$
I({\bm x})
\;=\;
\mathrm{tr}_\sigma \Big (\prod_{i=1}^d {\bm \sigma} \cdot \big (\widehat{{\bm x}_i - {\bm x}} - \widehat{{\bm x}_{i+1}-{\bm x}} \big ) \Big )
\;.
$$
Writing all terms, one first finds
$$
I({\bm x})
\; = \;\sum_{j=1}^{d+1} (-1)^j \;\mathrm{tr}_\sigma \left({\bm \sigma} \cdot (\widehat{{\bm x}_1- {\bm x}}) 
\cdots \underline{ {\bm \sigma} \cdot (\widehat{{\bm x}_j- {\bm x}}) }  \cdots ({\bm \sigma} \cdot \widehat{{\bm x}_{d+1}- {\bm x}}) \right) 
\;,
$$
where the underline designates a factor which is omitted. Thus with \eqref{Id1}
$$
I({\bm x})
\; =\;
(2\imath)^\frac{d-1}{2}\, d!\;\sum_{j=1}^{d+1} (-1)^j \;\mathrm{Vol}[{\bm 0}, \widehat{{\bm x}_1- {\bm x}}, \ldots ,\underline{ \widehat{{\bm x}_j- {\bm x}} },  \ldots ,\widehat{{\bm x}_{d+1}- {\bm x}}]
\;.
$$
The vertices can be re-ordered and it is convenient to translate the whole simplex, to continue:
$$
I({\bm x})
\; = \;
-(2\imath)^\frac{d-1}{2} \,d!\;
\sum_{j=1}^{d+1} \mathrm{Vol}[{\bm x}+\widehat{{\bm x}_1- {\bm x}}, \ldots, {\bm x},  \ldots ,{\bm x}+\widehat{{\bm x}_{d+1}- {\bm x}}]
\;,
$$
where the vertex ${\bm x}$ is located at the $j$-th position. At this point, it is useful to introduce the notations
$$
\mathfrak{S}_j({\bm x})
\;=\;
[{\bm x}+\widehat{{\bm x}_1- {\bm x}}, \ldots, {\bm x},  \ldots ,{\bm x}+\widehat{{\bm x}_{d+1}- {\bm x}}]
$$
and
$$
\mathfrak{S}
\;=\;
[{\bm x}_1, {\bm x}_2,\ldots, {\bm x}_{d+1}]
\;,
$$
where ${\bm x}_{d+1}=0$. The orientations of these simplexes are the same because each $\mathfrak{S}_j({\bm x})$ can be continuously deformed into $\mathfrak{S}$ without reducing the volume to zero. Now note that, for arbitrarily selected $j$, all vertices 
$$
{\bm x}+\widehat{{\bm x}_1- {\bm x}}, \ldots, \underline{{\bm x}+\widehat{{\bm x}_j- {\bm x}}},  \ldots ,{\bm x}+\widehat{{\bm x}_{d+1}- {\bm x}}
$$ 
of the simplex $\mathfrak{S}_j({\bm x})$ are located on the unit sphere centered at ${\bm x}$. As such, the facets of $\mathfrak{S}_j({\bm x})$ stemming from ${\bm x}$ define a $d$-dimenisonal sector of the unit ball. This sector will be denoted by $\mathfrak{B}_j({\bm x})$ and its orientation is taken to be the same as of $\mathfrak{S}_j({\bm x})$. The entire unit ball will be denoted by $\mathfrak{B}$ and its orientation will be taken to be the same as that of $\mathfrak{S}$. One key fact is that:
$$
\mathrm{Vol}( \mathfrak{S}_j({\bm x})) - \mathrm{Vol}( \mathfrak{B}_j({\bm x}))\; \sim \;|{\bm x}|^{-(d+1)} 
\qquad 
\mbox{\rm as}  \;\; |{\bm x}| \rightarrow \infty
\;.
$$
This enables one to break the integral into two terms:
\begin{align*}
\int_{\mathbb{R}^d} d{\bm x} \;I({\bm x})
& =\; - (2\imath)^\frac{d-1}{2}\, d!\; \sum_{j=1}^{d+1} \,\int _{\mathbb{R}^d} d{\bm x} \; [\mathrm{Vol}(\mathfrak{S}_j({\bm x})) - \mathrm{Vol}(\mathfrak{B}_j({\bm x})) ]   
\\
& \;\;\;\;\;\; - \;(2\imath)^\frac{d-1}{2}\, d!\; \int _{\mathbb{R}^d} d{\bm x}\; 
\sum_{j=1}^{d+1} \mathrm{Vol}(\mathfrak{B}_j({\bm x}))
\;.
\end{align*}
At this point let us note that
$$
\int _{\mathbb{R}^d} d{\bm x} \; [\mathrm{Vol} (\mathfrak{S}_j({\bm x})) - \mathrm{Vol} (\mathfrak{B}_j({\bm x})) ]=0
\;,
$$
which is a consequence of the odd-symmetry of the integrand relative to the inversion of ${\bm x}$ relative to the center of the facet ${\bm x}_1, \ldots,\underline{{\bm x}_j}, \ldots, {\bm x}_{d+1}$ of $\mathfrak{S}$. Furthermore:
$$
\sum_{j=1}^{d+1} \mathrm{Vol} ( \mathfrak{B}_j({\bm x}))
\;=\;
\left\{
\begin{array}{cc}
\mathrm{Vol}(\mathfrak{B})
& \mbox{if} \ {\bm x} \ \mbox{inside} \ \mathfrak{S}\;,  \\
0 & \;\;\mbox{if} \ {\bm x} \ \mbox{outside} \ \mathfrak{S}\;,
\end{array}
\right . 
$$
which follows because the solid angles corresponding to the facets of the simplex $\mathfrak{S}$, as viewed from ${\bm x}$, add up to the full solid angle if ${\bm x}$ is inside the simplex, and they add up to zero if ${\bm x}$ is outside the simplex. Hence
\begin{align*}
\int_{\mathbb{R}^d} d{\bm x}\;I({\bm x})
\; =\; 
-(2\imath)^\frac{d-1}{2}\, d!\; \mathrm{Vol}(\mathfrak{B}) |\mathrm{Vol}(\mathfrak{S})|
\;.
\end{align*}
Now the orientations of $\mathfrak{B}$ and $\mathfrak{S}$ are the same so that
$$
\mathrm{Vol}(\mathfrak{B}) \,|\mathrm{Vol}(\mathfrak{S})|
\;=\;
|\mathrm{Vol}(\mathfrak{B})|\,\mathrm{Vol}(\mathfrak{S})
\;=\;
\frac{2(2\pi)^{\frac{d-1}{2}}}{d!!}\;\frac{1}{d!}\;\mathrm{Det}({\bm x}_1,\ldots,{\bm x}_d)
\;,
$$
and the identity follows.\qed

\vspace{0.2cm}

The numerical invariant generated by the pairing of the Chern character with $K_1(\mathscr A)$ can be rightfully called the non-commutative odd Chern number. Indeed, note that the extension ${\rm Tr} \, \# \, \tau_d$ over $M_N(\mathbb C) \otimes \mathscr A$ is implemented by the substitution $\mathcal T \rightarrow {\rm tr}_N \otimes \mathcal T$ in the local formula. To simplify and be consistent with the notation, we will use the same symbol for both. Hence, for a unitary $u$ from $M_N(\mathbb C) \otimes \mathscr A$, the numerical invariant provided by the pairing  takes the form
\begin{equation}\label{NCOddChernNr}
\Ch_d(u) 
\;=\;
\frac{\imath(\imath \pi)^\frac{d-1}{2}}{d!!}  \sum_{\rho} (-1)^\rho \; \mathcal{T} \left(\prod_{i=1}^d u^\ast \partial_{\rho_i}u  \right)
\;.
\end{equation}
In the operator representation, $\Ch_d(u)$ takes exactly the form presented in \eqref{OddChernNr1bis}. Furthermore, when $\Omega$ reduces to just one point  and the magnetic field is turned off ({\it i.e.} for translationally invariant systems), \eqref{OddChernNr1bis} is nothing but the real-space representation \eqref{OddChernNr1} of the classical odd Chern number over the torus.

\vspace{.2cm}

When applied to the Fermi unitary element $u_F$, the theory developed so far will cover the physics of chiral unitary topological insulators, but only under the assumption of a spectral gap. Indeed, the standard Combes-Thomas estimate assures us that 
$$
|\langle \bm x |G(H_\omega)| \bm y \rangle| 
\;\leq \;
A e^{-\gamma |\bm x-\bm y|}
\;, 
\qquad 0 < A\,,\;\;\gamma <\infty \; ,
$$
for any $G$ holomorphic in a neighborhood of $\sigma(H)$. With the spectral gap assumption, the sign function of $H_\omega$ can be obtained using the holomorphic calculus, hence the assumptions of the theory are automatically satisfied (see {\it e.g.} \cite{DDS} for explicit estimates), and we can state at once:

\begin{corollary}
Let $h \in M_N(\mathbb C) \otimes \mathcal A_0$ be the element defining a covariant family of finite range Hamiltonians $\{H_\omega\}_{\omega\in\Omega}$ on $\mathbb C^{2N} \otimes \ell^2(\mathbb Z^d)$. Suppose that the zero energy lies in a spectral gap of $h$ and assume the chiral symmetry $JhJ= - h$. Then:
\begin{enumerate}[\rm (i)]

\item The element $q = {\rm sgn}(h)$ is in $M_{2N}(\mathbb C) \otimes \mathcal A$ and furthermore
$$
\sup_{\omega \in \Omega} |q(\omega,\bm x)| 
\;\leq\; A\, e^{-\gamma |\bm x|}
\;, 
\qquad 0 < A\;, \;\; 0<\gamma < \infty
\;.
$$
In particular, $q$ belongs to the smooth algebra $M_{2N}(\mathbb C) \otimes \mathscr A$. Consequently, $u_F \in M_{N}(\mathbb C) \otimes \mathscr A$ and  statement (iii) of Proposition~\ref{ChernCharacter} applies:
$$
{\rm Ch}_d(u_F)
\;= \;
\mathrm{Ind} \left( E_{\bm x_0} \pi_\omega (u_F) E_{\bm x_0}\right) \,\in\, \mathbb{Z} \;.
$$

\item Let $h_t$ be a deformation of the Hamiltonian in the sense that, for every $\bm x \in \mathcal R$, the hopping matrices $h_t(\omega,x)$ vary continuously in  $M_{2N}(\mathbb C) \otimes C(\Omega)$. Assume that the chiral symmetry holds and that the spectral gap remains open during the deformation. Then 
$$
\sup_{\omega \in \Omega}\, 
|q_t (\omega,\bm x)-q_{t'} (\omega,\bm x)| \leq A(t,t')e^{-\gamma |\bm x|}
\;, 
\qquad 0 < \gamma < \infty,
$$
with $A(t,t')$ continuous of both arguments and $A(t,t)=0$ for all $t$'s. In particular, $q_t = {\rm sgn}(h_t)$ varies continuously inside $M_{2N}(\mathbb C) \otimes \mathscr A$ and same can be said about $u_F(t)$. Consequently, statement (i) of Proposition~\ref{ChernCharacter} applies and ${\rm Ch}_d(u_F(t))$ remains constant and quantized during the deformation.

\end{enumerate}   
\end{corollary}

\section{The index theorem in the strong disorder regime}
\label{Sobolev} 

In the following, we leave the standard framework of non-commutative geometry and push the local index theorem over to a Sobolev space, using classical tools from functional analysis such as the Calderon-Fedosov formula\cite{Cal,Fed} as well as the invariance of the Fredholm index w.r.t. compact perturbations. We will work directly with elements from $M_N(\mathbb C) \otimes L^\infty(\mathcal A,\mathcal T)$ which will be represented on $\mathbb C^N \otimes \mathcal H$ using ${\rm id}\otimes \pi_\omega$. Also, the Dirac operator and its affiliated operators will be extended as $\bm 1 \otimes D$, {\it etc.}, but we will keep the same notations.

\begin{theorem}\label{IndexTh}
Let $u \in M_N(\mathbb C) \otimes L^\infty(\mathcal A,\mathcal T)$ be a unitary element which also belongs  to the Sobolev space $M_N(\mathbb C) \otimes \mathcal W'_{p,r}(\mathcal A,\mathbb P)$, with $p=r=d+1$. Then:
\begin{enumerate}[\rm (i)]

\item $\mathbb P$-almost surely, $E_{\bm x_0} \, \pi_\omega(u) \, E_{\bm x_0}$ is a Fredholm operator over $E_{\bm x_0}(\mathbb C^N \otimes \mathcal H)$. Its Fredholm index is $\mathbb P$-almost surely constant in $\omega \in \Omega$ and independent of $\bm x_0$. 

\item The following local index formula holds, $\mathbb P$-almost surely,
\begin{equation}\label{FinalIndex}
\mathrm{Ind} \Big ( E_{\bm x_0} \pi_\omega (u) E_{\bm x_0}\Big )
\;=\;
\Lambda_d  \sum_{\rho} (-1)^\rho \; \mathcal{T} \Big (\prod_{i=1}^d u^\ast \partial_{\rho_i} u  \Big )
\;.
\end{equation}

\item The $(d+1)$-linear functional
$$
(f_0,\ldots,f_d)\;\; \mapsto\;\; 
\sum_{\rho} (-1)^\rho\; \mathcal{T} \Big ( f_0 \prod_{i=1}^d \partial_{\rho_i}f_i  \Big )$$
is continuous on $M_N(\mathbb C) \otimes \mathcal W'_{p,r}(\mathcal A,\mathbb P)$.

\item The odd Chern number ${\rm Ch}_d(u_t)$ remains quantized and invariant for any deformation 
$$
t \;\;\mapsto\;\; u_t\in M_N(\mathbb C) \otimes \Big ( L^\infty(\mathcal A,\mathcal T)\cap \mathcal W'_{p,r}(\mathcal A,\mathbb P) \Big )
$$ 
which is continuous w.r.t. the norm $\| \cdot \|'_{p,r}$ (and not necessarilly w.r.t. $\|\cdot \|_{\infty}$).

\end{enumerate}

\end{theorem}

\proof (i) Let $K=E_{\bm x_0} \pi_\omega (u) E_{\bm x_0}$ defined over $E_{\bm x_0}(\mathbb C^N \otimes \mathcal H)$. The Claderon-Fedosov principle states that $K$ is Fredholm provided there is a positive integer $k$ such that $(E_{\bm x_0}-KK^\ast)^k$ and $(E_{\bm x_0}-K^\ast K)^k$ are trace class on $E_{\bm x_0} (\mathbb C^N \otimes \mathcal H)$. 
As it is well known, for $k=d+1$ the Calderon-Fedosov principle \cite{Cal,Fed} reduces precisely to condition \eqref{Summability1}. We will show that
\begin{equation}
\label{AvBound}
\int_\Omega \mathbb P(d\omega) \;  \mathrm{Tr}\left ( \big |\imath [F_{{\bm x}_0},\pi_\omega (f)] \big |^{d+1} \right)
\; \leq \;
A \ (\|f\|'_{p,r})^{d+1}
\;,
\end{equation}
for any $f\in M_N(\mathbb C) \otimes \mathcal W'_{p,r}(\mathcal A,\mathbb P)$. This ensures that the Calderon-Fedosov principle holds $\mathbb P$-almost surely. As in the proof of Proposition~\ref{Summability}, it is enough to consider only self-adjoint elements, $f = f^\ast$, in which case $\imath [F_{{\bm x}_0},\pi_\omega (f)]$ is self-adjoint and the absolute value in \eqref{AvBound} can be dropped (recall that $d+1$ is even). Finally, decomposing $f$ in a basis of $M_N(\mathbb C)$ and using Minkovski inequality for the Schatten norms, one can see that it is enough to take $f$ from $\mathcal W'_{p,r}(\mathcal A,\mathbb P)$ rather than from $M_N(\mathbb C) \otimes \mathcal W'_{p,r}(\mathcal A,\mathbb P)$. We will prove the estimate
\begin{equation}
\label{AvBoundP}
\int_\Omega \mathbb P(d\omega) \,  \sum_{\bm x' \in \mathbb Z^d} |\langle \bm x' \big |[F_{{\bm x}_0},\pi_\omega (f)]^{d+1} | \bm x \rangle |
\; < \;
A \,|\bm x+\bm x_0|^{-d-1} \ (\|f\|'_{p,r})^{d+1}
\;,
\end{equation}
which automatically implies \eqref{AvBound}. We denote the l.h.s. of \eqref{AvBoundP} by $W$ and start the calculation from Eq.~\eqref{KPower},
\begin{align*}
W 
\;\leq \;
 \frac{A}{ |\bm x + \bm x_0|^{d+1}}  \int_\Omega \mathbb P(d\omega) \sum_{\bm y_1,\ldots,\bm y_{d+1} \in \mathbb Z^d}  
 \prod_{i=1}^{d+1} \;(1+|\bm y_i|)^{d+1}\; |f(\omega_i, \bm y_i)|. 
\end{align*}
Since all the terms are positive, the sums and the integral can be interchanged and H\"older's inequality gives
\begin{align*}
W 
\;& \leq \;
 \frac{A}{ |\bm x + \bm x_0|^{d+1}} \sum_{\bm y_1,\ldots,\bm y_{d+1} \in \mathbb Z^d} \prod_{i=1}^{d+1} (1+|\bm y_i|)^{d+1}\; \left [  \int_\Omega \mathbb P(d\omega)  |f(\omega_i, \bm y_i)|^{d+1} \right ]^\frac{1}{d+1} \\
 &\; \leq \; \frac{A}{ |\bm x + \bm x_0|^{d+1}} \Big  (\sum_{\bm y \in \mathbb Z^d} (1+|\bm y|)^{d+1}\; \left [  \int_\Omega \mathbb P(d\omega)  |f(\omega, \bm y)|^{d+1} \right ]^\frac{1}{d+1} \Big )^{d+1},
\end{align*}
and \eqref{AvBoundP} follows. Next let us prove the $\mathbb P$-almost sure constancy of the index in $\omega$.  Since $\mathbb P$ is ergodic w.r.t. the lattice shifts, it is sufficient to check constancy along every orbit, namely, compare the indices of $E_{{\bm x}_0}\pi_\omega(u) E_{{\bm x}_0}+(\bm 1 -E_{{\bm x}_0})$ and $E_{{\bm x}_0} \pi_{\action_{\bm a}\omega}(u) E_{{\bm x}_0}+(\bm 1 -E_{{\bm x}_0})$ on the Hilbert space $\mathbb C^N\otimes\mathcal H$ for arbitrary ${\bm a}\in{\mathbb Z}^d$. Since the index is invariant to conjugations with unitaries, one only needs to check equality of the indices of $E_{{\bm x}_0}\pi_\omega(u) E_{{\bm x}_0}+(\bm 1 -E_{{\bm x}_0})$ and $E_{{\bm a}+{\bm x}_0}\pi_\omega(u) E_{{\bm a}+{\bm x}_0}+(\bm 1 -E_{{\bm a}+{\bm x}_0})$. But 
\begin{align*}
& \indent E_{{\bm a}+{\bm x}_0}\pi_\omega(u) E_{{\bm a}+{\bm x}_0} - E_{{\bm x}_0}\pi_\omega(u) E_{{\bm x}_0}  
\\
& \;\;\;\;\;\;\;=\;
\tfrac{1}{2}(F_{{\bm a}+{\bm x}_0}-F_{{\bm x}_0}) \pi_\omega(u) E_{{\bm a}+{\bm x}_0} +\tfrac{1}{2}E_{{\bm x}_0}\pi_\omega(u) (F_{{\bm a}+{\bm x}_0}-F_{{\bm x}_0} )
\;.
\end{align*}
The operator 
$$
F_{{\bm a}+{\bm x}_0}-F_{{\bm x}_0}
\;=\;
\bm 1 \otimes {\bm \sigma} \cdot (\widehat{{\bm a}+{\bm x}_0+{\bm X}} - \widehat{{\bm x}_0+{\bm X}})
$$
has the finitely degenerate singular values 
$$
|\widehat{{\bm a}+{\bm x}_0+{\bm x}}\, -\, \widehat{{\bm x}_0+{\bm x}}|
\;,
$$ 
which behave as $|{\bm a}+({\bm a}\cdot \hat {\bm x})\hat{\bm x}|/|{\bm x}|$ in the limit  $|{\bm x}|\rightarrow \infty$, hence this operator is compact. Similarly $E_{{\bm x}_0}-E_{{\bm a}+{\bm x}_0}$ is compact. The compact stability of the index now allows to conclude. The invariance in ${\bm x}_0$ follows by a similar argument.

\vspace{.1cm}

(ii) We have seen in (i) that $K=E_{\bm x_0}\pi_\omega(u)E_{\bm x_0}$ obeys $\mathbb P$-almost surely the Calderon-Fedosov principle with exponent $k=d+1$. Therefore, the Calderon-Fedosov formula allows to compute its index
$$
{\rm Ind}\big (E_{\bm x_0}\pi_\omega(u)E_{\bm x_0}\big ) 
\;=\;
{\rm Tr} \Big ( (E_{\bm x_0}-KK^\ast)^{d+1} \Big) - {\rm Tr} \Big ( (E_{\bm x_0}-K^\ast K)^{d+1} \Big )
\;.
$$
As well-known \cite{Con2}, this formula reduces to 
$$
{\rm Ind}\big (E_{\bm x_0}\pi_\omega(u)E_{\bm x_0}\big ) 
\;=\;
\lambda_d\, \mathrm{Tr}
\left(
F_{\bm x_0}[F_{\bm x_0},\pi_\omega (u^\ast)][F_{\bm x_0},\pi_\omega (u)] \cdots [F_{\bm x_0},\pi_\omega(u)]
\right)
\;,
$$
as in Proposition~\ref{ChernCharacter}. Since the index is $\mathbb P$-almost surely constant in $\omega$ and independent of $\bm x_0$, one is allowed to take the average of the r.h.s.. But this leads to \eqref{Cocycle}, with the appropriate arguments inserted. Hence, from here on, the calculation can proceed as in Theorem~\ref{LocalFormula}.

\vspace{.1cm}

(iii) It is enough to establish the continuity in norm $\| \cdot \|'_{d+1,d+1}$ for a dense subset of $M_N(\mathbb C) \otimes \mathcal W'_{d+1,d+1}(\mathcal A,\mathbb P)$, which can be conveniently chosen to be $\mathcal A_0$. Using H\"older's inequality \eqref{Holder}, the multi-linear functional is seen to be continuous in the standard Sobolev norm $\| \cdot \|_{d+1,1}$. Due to \eqref{NormInequality}, the functional is also continuous w.r.t. $\| \cdot \|'_{d+1,1}$. Lastly, due to \eqref{Comp1}, the functional is continuous in norm $\| \cdot \|'_{d+1,d+1}$.

\vspace{.1cm}

(iv) The $\mathbb P$-almost sure Fredholm index of $E_{\bm x_0}\pi_\omega(u)E_{\bm x_0}$ and the odd Chern number are linked by Eq.~\eqref{FinalIndex} and both sides of this equation must be used to establish the claim. Due to (iii) above,  the odd Chern number ${\rm Ch}_d(u_t)$ varies continuously with $t$, for any unitary homotopy $u_t$ in $M_N(\mathbb C) \otimes \mathcal W'_{p,r}(\mathcal A,\mathbb P)$. Now assume that this real number changes from one integer value to another as $t$ is varied. Due to continuity with $t$, then there must be at least one value of $t$ for which ${\rm Ch}_d(u_t)$ is not an integer. Now since the Fredholm index is {\it de facto} an integer, we have to conclude that actually there is not a single $\omega$ in the whole $\Omega$ for which \eqref{FinalIndex} holds at this $t$. But this will contradict the $\mathbb P$-almost sure character of equality \eqref{FinalIndex}. Hence, the starting assumption must be false and the conclusion is that ${\rm Ch}_d(u_t)$ stays pinned to a single integer value at all $t$'s. \qed 

\vspace{0.2cm}

When applied to the Fermi unitary element $u_F \in L^\infty(\mathcal A, \mathcal T)$, Theorem~\ref{IndexTh} covers the physics of chiral unitary topological insulators under the physically more interesting condition of a mobility gap, as the following proposition shows.

\begin{proposition}
Let $h \in M_N(\mathbb C) \otimes \mathcal A_0$ be the element defining a covariant family of finite range Hamiltonians $\{H_\omega\}_{\omega\in\Omega}$ on $\mathbb C^{2N} \otimes \ell^2(\mathbb Z^d)$. View $h$ as an element of the von Neumann algebra $L^\infty(\mathcal A,\mathcal T)$ (hence its functional calculus takes place there) and suppose that the zero energy lies in a mobility gap of $h$ where Aizenman-Molchanov bound \cite{Aizenmann1993uf} holds
\begin{equation}
\label{AM}
\int_\Omega \mathbb P (d\omega) \,|(h-E \pm \imath \, 0 ^+)^{-1}(\omega,\bm x)|^s
\; \leq\; 
C_s\, e^{-\gamma_s |{\bm x}|}
\;,
\end{equation}
for all $s \in (0,1)$ and strictly positive constants $A_s$ and $\gamma_s$. Assume also the chiral symmetry $JhJ= - h$. Then:
\begin{enumerate}[\rm (i)]

\item The element $q = {\rm sgn}(h) \in L^\infty(\mathcal A,\mathcal T)$ satisfies
$$
\int_\Omega \mathbb P(d\omega) \; |q(\omega,\bm x)| 
\;\leq \;
A \,e^{-\gamma_s |\bm x|}
\;.
$$
In particular, $q$ belongs to the Sobolev space $M_N(\mathbb C) \otimes \mathcal W'_{p,r}(\mathcal A,\mathbb P)$ with $p=r=d+1$. Consequently, $u_F$ belongs to $M_N(\mathbb C) \otimes \mathcal W'_{p,r}(\mathcal A,\mathbb P)$, too, hence statements (i) and (ii) of Theorem~\ref{IndexTh} apply for $u_F$.

\item Let $h_t$ be a deformation of the Hamiltonian in the sense that hopping matrices $h_t(\omega,\bm x)$, for every $\bm x \in \mathcal R$, vary continuously in  $M_{2N}(\mathbb C) \otimes C(\Omega)$. Assume that the chiral symmetry holds and that the mobility gap remains open such that \eqref{AM} holds uniformly with $t$. Then 
$$
\int_\Omega \mathbb P(d\omega) \, |q_t (\omega,\bm x)-q_{t'} (\omega,\bm x)| 
\;\leq \;A(t,t')\,e^{-\gamma |\bm x|}
\;,
$$
with $\gamma$ a strictly positive constant and $A(t,t')$ a continuous function of both arguments and such that $A(t,t) =0$. In particular, $q_t = {\rm sgn}(h_t)$ varies continuously inside $M_N(\mathbb C) \otimes \mathcal W'_{p,r}(\mathcal A,\mathbb P)$ and same can be said about $u_F(t)$. Consequently, statement (iv) of Theorem~\ref{IndexTh} applies for $u_F(t)$.

\end{enumerate}   
\end{proposition}

\proof (i) Since $q = \bm 1 - 2p_F$, it is enough to examine the Fermi projection, for which we use, like in \cite{Aizenman1998bf}, the identity 
\begin{equation}
\label{SignFunction}
p_F(\omega,\bm x)
\;=\;
\chi(h <0)
\;=\;
\oint_{\Gamma} \frac{dz}{2\pi \imath} 
\;(z-h)^{-1}({\omega,\bm x})
\; \in M_N(\mathbb C) \otimes L^\infty(\mathcal A,\mathcal T),
\end{equation}
where the contour $\Gamma$ encircles the negative spectrum and crosses the real axis (hence the essential spectrum) at $E_F = 0$. Let us add a few comments about this formula. First of all, it is well-known that, under the mobility gap assumption, $E_F = 0$ is $\mathbb P$-almost surely not an eigenvalue of $H_\omega$ and that the limits $\langle {\bm 0}|(E\pm\imath \,0^+-H_\omega)^{-1} |{\bm x}\rangle$ exist for Lebesgue almost all $E\in \mathbb R$. Thus $\mathbb P$-almost surely, the integral on the r.h.s. exists and actually defines a weak integral operator. The standard argument showing that Riesz projections are indeed orthogonal projections can be repeated and, furthermore, by using the spectral represenation of $H_\omega$ one deduces that this projection is indeed the Fermi projection. Now, to bound the r.h.s. of \eqref{SignFunction} let us recall the standard Combes-Thomas estimate
$$
|\langle {\bm 0}|(z-H_\omega)^{-1} |{\bm x}\rangle|
\;\leq\;
C\,|\Im m(z)|^{-1}\,e^{-C\,|\Im m(z)|\,|x|}
\;,
$$
for some finite constant $C$. Combining the disorder average of this with \eqref{AM}, one obtains a bound as \eqref{AM} uniformly along the entire $\Gamma$. Then, with the help of the universal bound $|(z-h)^{-1}(\omega,{\bm x})| \leq |{\rm Im} \, z|^{-1}$,
\begin{align*}
\int\limits_\Omega \mathbb P(d\omega) \; |p_F(\omega,\bm x)|
& \;\leq\;
\int\limits_{\Gamma} \frac{dz}{2\pi} \ |{\rm Im} \, z|^{s-1} \int\limits_\Omega \mathbb P(d\omega) 
\,|(z-h)^{-1}(\omega,{\bm x})|^s
 \;\leq\;
A\, e^{-\gamma_s |{\bm x}|}
\;.
\end{align*}

(ii) Let $\delta h = h_{t} -h_{t'}$. Since the deformation of the model is considered such that $h_t(\cdot, \bm x)$ for each $\bm x \in \mathcal R$ varies continuously with $t$ inside $M_N(\mathbb C) \otimes C(\Omega)$, the following function
$$
g(t,t')\; = \;\max_{\bm x \in \mathcal R} \; \sup_{\omega \in \Omega} |\delta h(\omega,\bm x)|.
$$
is continuous in both arguments and of course $g(t,t)=0$ for all $t's$. Now, Eq.~\eqref{SignFunction}, combined with the resolvent identity, shows
\begin{align*}
&\big |\big (p_F(t) -p_F(t')\big )(\omega, \bm x) \big |
\;\leq\;
\int_{\Gamma} \frac{|dz|}{2\pi} 
\,\Big | \Big ((z-h_{t})^{-1} \delta h (z-h_{t'})^{-1}\Big) (\omega, \bm x) \Big |
\\ 
 \leq\;
g & (t,t')\,\int_{\Gamma}\frac{|dz|}{|{\rm Im}\, z|^{1-s}}\,
\sum_{\bm x''-\bm x'\in \mathcal R} 
\big |(z-h_{t})^{-1}(\omega',\bm x')\big |^{\frac{s}{2}}
\,\big |(z-h_{t'})^{-1}(\omega'',\bm x - \bm x'')\big |^{\frac{s}{2}}
\;,
\end{align*}
where $\omega'$ and $\omega''$ are translates of $\omega$. Now the Cauchy-Schwarz inequality and the invariance of $\mathbb P$ against translations of $\omega$, together with the bound \eqref{AM} holding uniformly all along $\Gamma$, give
\begin{align*}
&
\int_\Omega \mathbb P(d\omega)\,
\big |\big (p_F(t) -p_F(t')\big )(\omega, \bm x) \big |
\\
& \;\;\;\;\leq\;
g(t,t') A \,\int_{\Gamma}\frac{|dz|}{|{\rm Im}\, z|^{1-s}}\,
\sum_{\bm x'-\bm x'' \in \mathcal R} 
e^{-\frac{1}{2}\gamma_s|{\bm x}'|}
\,
e^{-\frac{1}{2}\gamma_s|{\bm x}''-{\bm x}|}
\\
& \;\;\;
\;\leq\;
A(t,t')\,
e^{-\frac{1}{4}\gamma_s|{\bm x}|}
\;,
\end{align*}
with $A(t,t')$ as described in the statement. \qed


\end{document}